\documentclass[prb,twocolumn,showpacs,amsmath]{revtex4}
\usepackage{dcolumn}
\usepackage{bm}
\usepackage{graphicx}
\usepackage{color}
\usepackage{bm}
\begin{document}
\title{The interparticle interaction and crossover in critical lines on field-temperature plane in Pr$_{0.5}$Sr$_{0.5}$MnO$_{3}$ nanoparticles}
\author{A. K. Pramanik}
\altaffiliation[Present address: ]{Leibniz Institute for Solid State and Materials Research (IFW) Dresden, D-01171 Dresden, Germany.}
\author{A. Banerjee}
\affiliation{UGC-DAE Consortium for Scientific Research, University Campus, Khandwa Road, Indore-452001, M.P, India.}
\date{\today}
\begin{abstract}
The magnetic properties and the effects of interparticle interaction on it have been studied in nanoparticles of half doped Pr$_{0.5}$Sr$_{0.5}$MnO$_{3}$. Three samples consisting of nanoparticles of different average particle sizes are synthesized to render the variation in interparticle interaction. Though all the samples crystallize in the same structure to that of their bulk compound, the low temperature ferromagnetic-antiferromagnetic transition, which is present in bulk compound, is not evident in the nanoparticles. Linear as well as nonlinear ac susceptibility coupled with dc magnetic measurements have shown the superparamagnetic behavior of these nanoparticles where the blocking temperature increases with the increasing particle size. Presence of interparticle interaction is confirmed from the temperature variation of coercive field and the analysis of frequency dependent ac susceptibility. We have identified the nature of this interaction to be of dipolar type, and show that its strength decreases with the increasing particle size. The effect of this dipolar interaction on magnetic properties is intriguing as the compounds exhibit crossover from de Almeida-Thouless to Gabay-Toulouse like critical lines on field-temperature plane above their respective interaction field. In agreement with theoretical prediction, we infer that this crossover is induced by the unidirectional anisotropy arising from interparticle interaction, and this is confirmed from the presence of exchange bias phenomenon. 
\end{abstract}
\pacs{75.75.-c, 75.47.Lx, 75.40.Gb, 75.20.-g}
\maketitle
\section {Introduction}
The complex coupling between spin, charge, orbital and lattice in perovskite manganites around half doping gives rise to various exotic electro-magnetic properties. \cite{CNR-book,dago-book,tokura-rep} The of critical balance between the competing antiferromagnetic (AF) - insulating (I) and ferromagnetic (FM) - metallic (M) orders is found to be rather susceptible to small perturbations which can causing drastic effects.  \cite{ban, dag09}  In the recent time, it is shown that reduction in the particle size or dimensionality have rather significant effects on the physical properties. The charge ordering (CO), orbital ordering (OO) and low temperature ($T$) AF states are largely modified in the thin films of  Nd$_{0.5}$Sr$_{0.5}$MnO$_3$ and Pr$_{0.5}$Sr$_{0.5}$MnO$_{3}$. \cite{nsmo,uozu} In the parallel developments, it is demonstrated that in the nanoparticles, the notable bulk properties like: CO, OO, low temperature AF states are largely suppressed or weakened, and the system emerges to FM state. \cite{raoncmo,idas-psmo,tapati-lcmo, jirak} However, it was suggested by N$\acute{e}$el long back that small or finite-size particles of AF material should exhibit superparamagnetism (SPM) or weak FM properties. \cite{neel-AF}

The physical properties in nanoparticles are mainly modified due to (i) the finite size effect, and (ii) the surface disorder effect which is caused by defects, broken exchange bonds, fluctuations in number/separation of neighboring atoms at the surface. Usually, FM particles with reduced size behave as a single domain entity exhibiting SPM behavior where the thermal fluctuation overcome the anisotropy energy of individual particle causing the magnetization to flip from one to another easy direction. The situation becomes complicated if there exists interparticle interactions which induces collective behavior, and may modify the magnetic properties drastically. For instance, it is observed for ferrofluid that low-$T$ dynamics changes from SPM to spin glass (SG) like behavior with just increasing interaction among the particles. \cite{jonsson} Therefore, it is necessary to identify the nature as well as the extent of this interaction to characterize the magnetic properties of the collection or conglomerate of nanoparticles. Commonly, the inter-particle interaction is achieved or modulated with the variation of volume concentration of particles, however, the size variation of particles offer another avenue in this regard which does not require the intervening materials unlike the former case, and is uncommon in literature. 

Here we present a detailed study of the magnetic properties and their evolution with the inter-particle interaction in nanoparticles of half doped manganite, Pr$_{0.5}$Sr$_{0.5}$MnO$_{3}$ (PSMO). In this study, we have modulated the inter-particle interaction by means of modifying the size of the particles. Nanoparticles of three different average sizes (in the range of 15-27 nm) are synthesized which have identical room temperature crystal symmetry as their bulk counterpart. However, contrary to the bulk PSMO, these nanoparticles do not show the low temperature FM to AF transition.\cite{tomioka, ashim}  Using linear and non-linear ac susceptibility coupled with dc magnetization, we have confirm that these nanoparticles show SPM behavior where the blocking temperature ($T_B$) increases with the particle size.  Analysis of frequency dependent peak in ac susceptibility and temperature variation of coercive field indicate the presence of interparticle interaction. The nature of the interaction is found to be of dipolar type, and its strength decreases with increasing particle size. It is rather significant that  with increasing the field ($H$), all the nanoparticles exhibit a crossover in critical lines from the de Almeida-Thouless to the Gabay-Toulouse like behavior around the respective dipolar interaction field. Following the theoretical prediction, we believe that this crossover is induced by unidirectional anisotropy which, in the present systems arises from the interparticle interaction and is confirmed from the presence of exchange bias effect.

\maketitle\section{Experimental Details}
Nanocrystalline samples of PSMO are prepared, using Pr$_6$O$_{11}$, SrCO$_3$ and Mn(CH$_3$COO)$_2$ with purity of 99.99\% or better, by chemical (pyrophoric) method. \cite{pramanik} First, the proportional amount of ingredients are dissolved in dilute HNO$_3$. Triethanolamine (TEA) is mixed with this aqueous solution of the ingredients in the metal ions to TEA ratio as (Pr,Sr):Mn:TEA = 1:1:4, and evaporated around 200$^0$C. At the end, it yields black precursor powders which are heated at 600, 650 and 700$^0$C for 5 hours to get the powders of different nanosize  samples which are designated as N600, N650 and N700, respectively. Room temperature x-ray diffraction (XRD) patterns are collected with Rigaku Dmax 300 diffractometer attached to 18 kW rotating anode (Cu) source. Rietveld refinement \cite{young} of XRD patterns show that samples are in single phase, and they crystallize in tetragonal structure with \textsl{I4/mcm} symmetry similar to their bulk counterpart. Fig. 1a presents XRD data along with the Rietveld fitting for the N650 sample. Average particle size ($D_{XRD}$) has been estimated from the XRD plot using Scherrer formula. \cite{cullityxrd} The calculated particle sizes are given in Table 1. Structural parameters found from the Rietveld refinements are also given in Table 1. There is no major structural modification in nanometric samples compared to the bulk. \cite{ashim} Extreme right column of Table 1 presents the percentage change in the related parameters for all the samples, showing that no major structural modification occurs with the size variation of particles.

\begin{table}
\caption{\label{tab:table 1} Structural parameters are determined from the Rietveld refinement of the powder XRD patterns for nanocrystalline Pr$_{0.5}$Sr$_{0.5}$MnO$_{3}$. Here O1 is the apical and O2 is the equatorial oxygen in perovskite structure. $D_{XRD}$ and $D_{TEM}$ are the average particle sizes calculated from the XRD and TEM measurements. $T_B$ is the blocking temperature obtained from ac susceptibility. The extreme right column shows the percentage change ($\Delta\%$) in structural parameters for all the nanoparticles.}
\begin{ruledtabular}
\begin{tabular}{cccccc}
Samples &N600 &N650 &N700 &$\Delta\%$\\
\hline
$a$ (\AA) &5.4270(7) &5.4259(5) &5.4255(5) &-0.02\\
$c$ (\AA) &7.7267(17) &7.7226(12) &7.7213(12) &-0.06\\
$V$ (\AA${^3}$) &227.57(6) &227.35(5) &227.29(5) &-0.12\\
Mn-O1 (\AA)  &1.9317 &1.9306 &1.9303 &-0.07\\
Mn-O1-Mn  &180\raisebox{1ex}{\scriptsize o} &180\raisebox{1ex}{\scriptsize o} &180\raisebox{1ex}{\scriptsize o} &0.0\\
Mn-O2 (\AA)  &1.9337 &1.9278 &1.9303 &-0.25\\
Mn-O2-Mn  &165.7\raisebox{1ex}{\scriptsize o} &168.6\raisebox{1ex}{\scriptsize o} &167.1\raisebox{1ex}{\scriptsize o} &+1.7\\
$D_{XRD}$ (nm) &15.7 &17.3 &19.1 &-\\
$D_{TEM}$ (nm) &15.7 &19.2 &26.6 &-\\
$T_B$ (K) &209.8 &240.3 &253.3 &-
\end{tabular}
\end{ruledtabular}
\end{table} 

Samples are also characterized through transmission electron microscope (TEM) (model: TECNAI G2-20FEI). Fig. 1b shows TEM bright field image of N650 sample. Histogram for particle size distribution is obtained after analyzing several TEM images. Such histogram for N650 sample, presented in Fig. 1c, shows that particles are polydisperse in nature. Mean particle size ($D_{TEM}$) for all the samples have been obtained from the Gaussian fitting of figure similar to Fig. 1c and values are given in Table 1. The values of $D_{XRD}$ and $D_{TEM}$ match reasonably well except for N700 sample. This mismatch in sizes for N700 may be arising from conglomeration of particles because of higher temperature processing. Selected area electron diffraction (SAED) graph has been given in Fig. 1d for N650 sample. The figure shows well distinct spotty concentric ring indicating crystalline nanoparticles. The similar figure for N600 sample is given in Ref. \onlinecite{NP-conf}. The oxygen stoichiometry or Mn$^{3+}$/Mn$^{4+}$ ratio of 1:1 is ensured from Iodometric Redox titration.  Low field ac susceptibility measurements are done with home made ac susceptometer. \cite{ashnarsi} DC magnetization have been measured with homemade vibrating sample magnetometer (VSM) \cite{krsnarsi} and commercial 14 Tesla VSM (PPMS) made by Quantum Design. 
    
\maketitle\section{Results and discussions}
\maketitle\subsection{Superparamagnetic behavior of nanoparticles}
Fig. 2 presents the temperature variation of dc magnetization ($M$) measured in 100 Oe following zero field cooled (ZFC) and field cooled warming (FCW) protocols. ZFC magnetization shows broad peak at characteristic temperature ($T_B$) which is unlike its bulk compound where the sample exhibits clear PM-FM and then FM-AF transition on cooling from room temperature. \cite{tomioka, psmo-crit} The observed peak which is typical feature of metastable magnetic systems like SPM particles or SG system, shifts to higher $T$ with increasing particle size. It is clear in figure that ZFC and FCW branches of magnetization split at temperature ($T_{irr}$) higher than $T_B$, and $T_{irr}$ reduces with increasing the field (not shown). The higher value of $T_{irr}$ than $T_B$ indicates the presence of larger size spin clusters which are ordered at higher temperature. Contrary to the bulk PSMO, the first order FM-AF transition at lower temperature as well as the associated hysteresis between FCW and the field cooled cooling (FCC) magnetization, down to 2K, is not evident in these nanoparticles. Similar phenomenon, however, is observed in various nanocrystalline half doped manganites. \cite{raoncmo,idas-psmo,tapati-lcmo} From the Curie-Weiss fitting of high-T magnetization data [$(\chi = M/H)^{-1}$ vs $T$], the effective PM moment ($\mu_{eff}$) has been calculated in terms of Bohr magneton ($\mu_B$) per formula unit (f.u.). The obtained values are 3.62, 3.70 and 4.38 $\mu_B$/f.u. for N600, N650 and N700 compound, respectively. It can be noted that $\mu_{eff}$ for smaller particles are lower than the expected spin only value [$\mu_{eff}$ = g$\sqrt{S(S + 1)}$, where $S$ is the total spin] which is 4.38 $\mu_B$/f.u. for PSMO. It is noteworthy that with increasing particle size, $\mu_{eff}$ approaches the expected value.

To understand the low-$T$ magnetic state in these nanoparticles we have collected high field (120 kOe) $M$ vs $H$ plots at 2 K (Fig. 3a). Unlike the bulk PSMO which at low $T$ (within AF state) exhibits field induced jump in $M(H)$ and attains the saturation moment (3.5 $\mu_B$/f.u.),\cite{ashim,ashimjpcm} the nature of $M(H)$ plots for nanoparticles resemble qualitatively with FM systems. However, even in high field, magnetization does not saturate but show monotonous increase. The magnetic moments at 2 K and 120 kOe ($\mu$) are 1.38, 1.40 and 1.94 $\mu_B$/f.u.  for N600, N650 and N700 sample respectively. These values are much lower than the saturation moment of bulk PSMO (3.5 $\mu_B$/f.u.). Nonetheless, the estimated moments imply that the increased surface to volume ratio at smaller particles creates higher amount of magnetically dead layer at the surface, thus it reduces the average moment systematically. In Fig. 3b, $M(H)$ data are plotted in the form of $M^2$ vs $H/M$ which is known as Arrott plot.\cite{arrott} Positive intercept on the $M^2$ axis from the extrapolation of high field data in Arrott plot implies the presence of spontaneous magnetization in system. The spontaneous magnetization calculated from these intercepts (Fig. 3b) are 0.943, 1.003 and 1.367 $\mu_B$/f.u. for N600, N650 and N700 sample respectively. This clearly indicates that the ground state of these nanoparticles are FM contrary to their bulk counterpart even though there is no change in the structural symmetry.  

To confirm the magnetic state of these nanoparticles we have measured ac susceptibility ($\chi$) which probes the dynamics of the spin system in very low field. In general, magnetization can be expressed in terms of measuring field as:
\begin{eqnarray}
	M = M_0 + \chi_1H + \chi_2H^2 + \chi_3H^3 + \chi_4H^4 + ....
\end{eqnarray}
where M$_0$ is the spontaneous magnetization, $\chi_1$ is the linear and $\chi_2$, $\chi_3$, $\chi_4$ are the nonlinear susceptibilities. Fig. 4a presents temperature variation in real part of $\chi_1$ ($\chi_1^R$) measured in ac field 3 Oe and frequency 731 Hz. The feature of $\chi_1^R$ is similar to its dc magnetization counterpart presented in Fig. 2. The $T_B$ calculated from low field ac-$\chi$ are mentioned in Table 1. 

It is rather nontrivial issue to distinguish between the SPM and SG as both share similar kind of experimental features. To do this, we have used nonlinear ac susceptibilities which are shown to be an useful experimental probes to separate out different magnetic states. \cite{ashim,ashna,akm,ranganathan,bitoh,sunilnsus} In Fig. 4b, we have plotted temperature dependence of second order ($\chi_2$) ac-$\chi$ measured in ac field 3 Oe and frequency 731 Hz. Usually, $\chi_2$ appears due to the presence of symmetry breaking field which originates either from the spontaneous magnetization or the superimposed dc magnetic field. \cite{ashim,akm,ranganathan} As the canonical SG does not hold spontaneous magnetization, so $\chi_2$ should be absent without dc field. The presence of finite $\chi_2$ in Fig. 4b confirms the presence of FM interaction within these nanoparticles and discards the possibility of canonical SG like phase. However, it is important to identify if the broad peak in ZFC magnetization and in $\chi_1^R$ can be attributed to cluster-glass like phase or SPM. This is achieved through the measurement of 3rd order ac-$\chi$ [$\chi_3$($T$)] which show  broad negative peak around $T_B$ for all the samples (not shown). The SPM like feature in the present nanoparticles is confirmed from the noncritical behavior of the $\chi_3$ around the peak temperature. It has been unambiguously shown that, for SG like system, $\chi_3$ diverges as $H$ $\rightarrow$ 0 around the transition temperature.\cite{ashna,suzuki} In Fig. 5, we have plotted the maximum value of $\chi_3$ normalized by that at $H$ = 20 Oe as a function of measuring field. As evident in figure, $\chi_3$ is not divergent as $H$ $\rightarrow$ 0. This analysis shows conclusively that PSMO nanoparticles do not show SG like coperative freezing but thermal blocking of the magnetic entities. 

Further, the SPM behavior in these nanoparticles is also substantiated from the Wohlfarth's model.\cite{wohl,bitoh}  The magnetization for an assembly of particles is given as: 
\begin{eqnarray}
	M = n\left\langle \mu \right\rangle L (\left\langle \mu \right\rangle H/k_BT)
\end{eqnarray}
where $n$ is the number of particles per unit volume, $\left\langle \mu \right\rangle$ is the average moment of magnetic particle, $k_B$ is the Boltzmann constant and $L(x)$ is the Langevin function. After the expansion of $L(x)$, $\chi_1$ and $\chi_3$ above the blocking temperature ($T_B$) can be expressed as:
\begin{eqnarray}
	\chi_1 = n \left\langle \mu \right\rangle^2/3k_BT = P_1/T
\end{eqnarray}
\begin{eqnarray}
	\chi_3 = (n\left\langle \mu \right\rangle/45)(\left\langle \mu \right\rangle/k_BT)^3 = P_3/T^3
\end{eqnarray}
Therefore, $\chi_1$ and $\chi_3$ above $T_B$ for SPM will vary as $T^{-1}$ and $T^{-3}$, respectively. The main panels of Figs. 6a and 6b show this variation of $\chi_1^R$ and $\chi_3^R$ following Eqs. 3 and 4, respectively for N650 and N700 samples. In the inset of both the figures we show the same plot for N600 sample. This further corroborates the earlier analysis that peak in $\chi_1$ arises due to the blocking of magnetic clusters in the present study. From the ratio of $P_3$ and $P_1$, $\left\langle \mu \right\rangle$ is estimated which are around 14.5 $\times$ $10^4$, 22.9 $\times$ $10^4$ and 26.1 $\times$ $10^4$ $\mu_B$ for N600, N650 and N700 sample, respectively. This large value of $\left\langle \mu \right\rangle$ is consistent with the SPM clusters as it consists of large number of spins, whereas for PM, $\left\langle \mu \right\rangle$ signifies only ionic moments and is limited to few $\mu_B$. Assuming that clusters are spherical, effective sizes of them are calculated from the values of $\left\langle \mu \right\rangle$ derived from above analysis. These sizes are found to be around 16.5, 19.2 and 20.9 nm for N600, N650 and N700, respectively. It is noteworthy that effective size of clusters calculated from the magnetic measurement are close to the same obtained from XRD and TEM measurements (Table 1). This consistency of particle size calculated from various methods proves the unambiguity in present analysis.

\maketitle\subsection{Interaction among the nanoparticles}

After having confirmed the SPM nature of these nanoparticles, we attempt to establish the existence of interparticle interaction. As a first step, interparticle interaction is identified from the temperature dependence of coercive field ($H_C$) estimated from the $M(H)$ plot for all the samples. For the noninteracting SPM particles, temperature variation in $H_C$ is expressed as: \cite{cullity}
\begin{eqnarray}
	H_C = H_{C,0}\left[\left(1 - T/T_B\right)^{1/2}\right]
\end{eqnarray}
where $H_{C,0}$ is the value of $H_C$ when $T$ $\rightarrow$ 0. Thus, the measured $H_C$ as a function of $T^{1/2}$ would be linear for noninteracting particle systems. However, nonlinear behavior of $H_C$ vs $T^{1/2}$ in Fig. 7 arises due to the interaction among the particles. It is also evident in the figure that with increasing particle size nonlinearity of plot is reduced, thus probably indicate that the strength of interaction reduces with the increasing particle size.

Now we attempt to reestablish the presence of  interaction among the particles from the frequency ($f$) dependent shift in $T_B$ in ac-$\chi$($T$) measurements.   It can be mentioned that the variation of $f$ in ac-$\chi$ renders the variation in probe time $\tau_{P}$ ($\propto$ $f^{-1}$) which allows to probe the relaxation of particles in different time windows. It is observed in Fig. 8 that $\chi_1^R$ for N600 sample decreases and $T_B$ shifts toward higher $T$ with increasing $f$. However, the shift of $T_B$ with the $f$ in present study is relatively small. It can be mentioned that simple, non-interacting SPM particles show large $f$ dependence of $T_B$ whereas for SG or interacting SPM particles, $T_B$ is less $f$ dependent. \cite{tholence} The $f$ dependence of $T_B$ is normally quantified with the following empirical relation:

\begin{eqnarray}
 \Phi = \frac{\Delta T_B}{T_B \Delta \log_{10}(f)}	
\end{eqnarray}
where $\Delta$ is the difference in related parameters. The value of $\Phi$ for N600 sample is found to be around 0.002. The other two samples also show similar values of $\Phi$. The experimentally determined values of $\Phi$ for SPM particles are in the range of 0.1 - 0.13, whereas much lower values (0.005 - 0.05) are observed for canonical SG systems as well as interacting particle system. \cite{goya,toro} Therefore, based on the value of $\Phi$ it is often difficult to distinguish between SPM and SG experimentally. However, the calculated $\Phi$ hints toward the presence of interaction among these particles.

To unambiguously identify the presence of interparticle interaction, we follow a rigorous systematic method where the $f$ dependence of $T_B$ has been analyzed with different phenomenological models. According to N$\acute{e}$el-Arrhennius law, relaxation time ($\tau$) for an assembly of noninteracting SPM particles behaves like:\cite{youngsg}
\begin{eqnarray}
	\tau = \tau_0 \exp \left(\frac{E_a}{k_BT}\right)
\end{eqnarray}
$E_a$ is the anisotropy energy barrier equal to KV, where K is the anisotropy energy constant and V is the volume of the clusters. For the SPM relaxation, the prefactor $\tau_0$ is in the range of $10^{-8}$ - $10^{-13}$ s. \cite{dormann-jmmm} Inset of Fig. 8 shows the straight line fitting of Eq. 6 for N600 sample. Though fit is apparently good but the obtained values are quite unphysical ($\tau_0$ $\approx 10^{-522}$ and $E_a/k_B$ = 253615) which clearly indicates that dynamics of these nanoparticles can not be explained with noninteracting particle model. Thus, we have tried to analyze the $f$ dependent shift of $T_B$ using Vogel-Fulcher law: \cite{youngsg}
\begin{eqnarray}
	\tau = \tau_0 \exp \left[\frac{E_a}{k_B(T - T_0)}\right]
\end{eqnarray}
The term $T_0$ is the characteristic temperature ($0 < T_0 < T_B$) which accounts the interaction among the particles. In this case, best fitting of data for N600 has been shown in Fig. 9a. The obtained fitted parameters are quite reasonable with the values $E_a$/$k_B$ = 25.27 $\pm$ 3.34 K, $T_0$ = 205.15 K and $\tau_0$ = 1.76 $\times$ $10^{-8}$ s. Such higher value of $\tau_0$ has been observed for interacting clusters. \cite{djurberg} This fitting confirms the interacting nature of these nanoparticles, and is in conformity with the recent observations like slow magnetic relaxation with logarithmic time dependence and memory effects.\cite{NP-conf}

It is rather significant and intriguing that the $f$ dependent shift of $T_B$ can be fitted to the scaling law as well, which is used to characterize the phase transition in SG, though the present system are proved to be almost tailor-made small particles, which behave like SPM beyond doubt. The scaling hypothesis assumes that the relaxation time ($\tau$) is related to the correlation length ($\xi$) near to the transition temperature ($T_g$). As $\xi$ diverges at $T_g$, relaxation time obeys the following empirical relation: \cite{youngsg}
\begin{eqnarray}
	\tau = \tau_0 \left(T/T_g - 1\right)^{-z\nu}
\end{eqnarray}
where $z$ is the dynamical scaling exponent and $\nu$ is the critical exponent related to $\xi$. Fig. 9b shows the best fit of Eq. 8 for N600 sample and the obtained parameters are $\tau_0$ = 3.39 $\times$ $10^{-14}$ s, $T_g$ = 206.44 K and z$\nu$ = 4.52. Usually, the exponent $z\nu$ shows large variation for different kind of systems. \cite{souletie} Although, the obtained $z\nu$ is close to the theoretically predicted value (4) for 3D Ising model, \cite{binder} and the experimentally calculated value (5.5) for canonical SG CuMn$_{4.6\%}$. \cite{souletie} However, $\tau_0$ shows large disagreement between Vogel-Fulcher law and scaling law fitting. Further, $\tau_0$ obtained in later case is orders of magnitude less than that usually seen for SG ($10^{-11}$ - $10^{-12}$ s). This inconsistency of $\tau_0$ with that for SG is rather significant, which discards the possibility of SG like freezing in these nanoparticles. \cite{youngsg} It is fact that the significant presence of interparticle interaction induces the collective behavior and glassy dynamics even in the tailor made SPM system. Since the response of such systems is similar to SG, they are termed as super spin glass (SSG).\cite{jonsson,djurberg-PRB,chen,suzuki-ssg} Basically, SSG dynamics differ from the SPM one in terms of critical slowing down of relaxation time. Similar critically slow dynamics resembling SG like response is also shown by magnetic clusters forming in a matrix out of composition fluctuation or atomic segregation which is called cluster glass (CG).\cite{gruzalski} However, it remains an experimentally challenging task to discern between critically slow dynamics of SSG and CG with the slow but non-critical dynamics of SPM system. The high value of $\tau_0$ and $T_0$ obtained from the Vogel-Fulcher law fitting unambiguously imply the presence of interparticle interaction in present nanoparticles,\cite{tholence} however, the obtained $\tau_0$ from the dynamical scaling hypothesis fitting is a few orders lower than the value for SSG or CG ($\tau_0 = 10^{-9} - 10^{-6}$).\cite{jonsson,djurberg-PRB,chen,suzuki-ssg} Nonetheless, the fitting of scaling law for our data is quite intriguing indicating some sort glassiness in the system, however, the overall results do not comply with the SSG behavior. This indicates that a careful exercise is needed to make a proper distinction between the SSG/CG like phases with that of SPM system from the dynamical behavior.
    
\maketitle\subsection{Nature of interparticle interaction and its consequences on magnetic properties}
To identify the nature of the interparticle interaction and its consequences on the magnetic properties, we have measured the remnant magnetization following isothermal remnant magnetization (IRM) and dc demagnetization (DCD) protocols. \cite{butera} For the IRM measurement, we cooled the sample in zero field from the room temperature to 150 K and applied a field ($H_a$) isothermally for 10 s. Then the $H_a$ is reduced to zero and remnant magnetization is measured. We heated the sample back to room temperature and repeated this experiment for different $H_a$ at the same temperature, each time increasing $H_a$ in steps of 50 Oe up to maximum of 1000 Oe. The measured IRM as function of $H_a$ for N600 sample is depicted in Fig. 10a. In DCD measurement, the sample is zero field cooled from room temperature to 150 K, and magnetized to saturated state with applying +1000 Oe for 10 s. The applied positive field is made zero, and isothermally a negative field ($-H_a$) is applied for 10 s. Then, $-H_a$ is switched off and remnant magnetization is measured. We repeated this experiment similar to IRM measurement for different $-H_a$ up to -1000 Oe in step of -50 Oe. The measured DCD as function of $|H_a|$ for N600 sample is depicted in Fig. 10a. It can be mentioned that IRM and DCD originate from the virgin and saturated state, respectively.
 
To characterize as well as quantify the magnetic interaction, the parameter $\delta M$ is shown to be very useful which is defined as: \cite{butera,wohlfarth}
\begin{eqnarray}
	\delta M = m_{DCD} - \left(1 - 2m_{IRM}\right)
\end{eqnarray}
where m is the normalized remnant magnetization with respect to the remnant magnetization measured after the sample is magnetized to saturated state. For noninteracting particles $\delta M$ is zero. \cite{wohlfarth} However, finite interactions among the particles lead to deviation of $\delta M$ from zero, i.e., positive $\delta M$ is due to the interactions which favor magnetization whereas negative $\delta M$ arises from the interactions which cause demagnetization in the system. \cite{butera} Usually, the negative $\delta M$ implies the dipolar interaction and the field where minimum in $\delta M$ occurs quantifies the strength of interaction. \cite{curiale} We have plotted $\delta M$ as a function of $|H_a|$ for N600 sample in Fig. 10b. As evident in figure, $\delta M$ is negative in low $H_a$, showing minimum around 150 Oe. In higher $H_a$ around 522 Oe, $\delta M$ shows crossover to positive value. We can infer from this results that interparticle interaction is of \textit{dipolar} type, having magnitude roughly of 150 Oe. At high field, interparticle interaction is dominated by the applied field yielding positive $\delta M$. For the same measurements on N700 compound, $\delta M$ shows negative minimum around 100 Oe, and crossover to positive value around 220 Oe. These results indicate that with the increasing particle size, strength of dipolar interaction reduces. The dipolar interaction between two magnetic dipoles $\vec{\mu_i}$ and $\vec{\mu_j}$ can be estimated as:\cite{youngsg}

\begin{eqnarray}
 E_{d} = \frac{1}{r_{ij}^3}\left[\vec{\mu_i} . \vec{\mu_j} - 3 \left(\vec{\mu_i} . \hat{r}_{ij}\right)\left(\vec{\mu_j} . \hat{r}_{ij}\right)\right]    	
\end{eqnarray}
 
where $r_{ij}$ is the center to center distance between the dipoles. Eq. 11 gives the form of long-range dipolar interaction energy between the two magnetic diploes of moment $\left\langle \mu \right\rangle$ separated by distance $r_{ij}$. For a collection of small particles, it has to be summed over all the particles. The present system may be considered as an assembly of spherical nanoparticles of radius $R$, placed adjacent to each other in a regular array. In this case, the separation between the adjacent magnetic moments, $r_{ij}$, is about 2$R$. If we consider the magnetic moment $\left\langle \mu \right\rangle$ of each particle is roughly proportional to its volume, then according to Eq. 11, dipolar interaction energy ($E_{d}$) between the two adjacent particles varies approximately as $R^3$. Since for a collection of particles within a given volume, the number of particles varies as inverse of $R^3$, the total interaction energy which is the summation over the entire individual $E_{d}$s, becomes almost independent of the particle size. However, in real situation, as the average particle size decreases, the packing fraction increases in such densely packed assembly of nanoparticles having a distribution in particle size. Thus decrease in the size will result in the increase in the number of particles in the same volume, much more than what is expected for a regular array where it increases as 1/$R^3$. Moreover, for such dense packing, reduction in the size of the particles will also reduce the average distance between them, $r_{ij}$, faster than a linear function of $R$. Both this factors will contribute to increase the total dipolar interaction energy in the system with the decrease in particle size. The evidence of such increase in dipolar interaction energy with the decrease in the particle size in the present system can be found in the Fig. 7 and the discussions of sections B and C.

The dipolar interaction has significant influences on the magnetic properties as it modifies the otherwise uniaxial anisotropy energy barrier of individual nanoparticle. The most notable effects of this interaction are the evolution of glassy dynamics,\cite{jonsson} modification in $T_B$,\cite{jonsson,morup} etc. Here we have looked into the effect of dipolar interaction on the field dependence of $T_B$ by measuring ZFC magnetization in different fields. In the experimental scenario, $T_B$ corresponds to the temperature where $\tau_P$ equals to the average $\tau$ of the system. Since with the variation of applied field $T_B$ varies, thus one can construct a line of constant $\tau$ on the plane spanned by $H$ and $T$. The existence of such line is predicted in the mean-field theoretical model for SG. To be specific, this model assumes phase transition in SG, and considers that large field will destroy the frozen spin state. \cite{youngsg} Therefore, in presence of field, the critical lines are predicted on the H-T plane which mark the phase transition. The first one is the de Almeida-Thouless (AT) line occurring in anisotropic Ising SG and behaves as $T_B$($H$) $\propto$ $H^{2/3}$. The second one is the Gabay-Toulouse (GT) line valid for isotropic Heisenberg case, and shows functional form $T_B$($H$) $\propto$ $H^{2}$.\cite{youngsg} However, later numerical calculation showed that such lines are not unique for SG, and can even exist in case of relaxation of (interactive) SPM particles with the crossover from low-$H$ GT to high-$H$ AT behavior.\cite{wenger} Our results are plotted in Fig. 11 where we show that variation of $T_B$ with field follows AT line in low field for all the samples. However, with the increase in field we find the deviation from AT line as marked by vertical arrows in the figure. Remarkably, in high field we find the variation of $T_B$(H) agrees with the GT line. We have observed the similar behavior for all the samples and in inset of Fig. 11 we have plotted such high field behavior for N600 sample. The fields ($H_{cr}$) where we find crossover from AT to GT behavior are approximately 250, 200 and 100 Oe for N600, N650 and N700 sample, respectively. It is rather significant that the crossover fields from AT to GT like behavior are close to the dipolar interaction fields found from the remnant magnetization measurements mentioned above (Fig. 10). This crossover in critical lines around the dipolar interaction field for these nanoparticles is quite noteworthy.

The above experimental results in Fig. 11 has the similarity with the theoretical calculation by Kotliar and Sompolinsky\cite{kotliar} who predicted that in presence of unidirectional random anisotropy (i.e., Dzyaloshinsky-Moriya type interaction) the critical behavior for SG in fields lower than the anisotropy is close to Ising-like following AT line, and crosses over to Heisenberg behavior in high fields. In fact, similar anisotropy induced crossover in critical lines is experimentally observed in Au doped classical CuMn SG system as well as in superconducting vortex-glass system.\cite{court,rodri} These results allow us to draw an analogy with the present PSMO nanoparticles where the interparticle interaction acts as an unidirectional anisotropy leading to crossover from AT to GT like behavior at field around the interaction field. Indeed, dipolar interaction (Eq. 11) is anisotropic which introduces angular dependence of spin ordering in system.\cite{youngsg}

To confirm the unidirectional nature of the anisotropy arising out of interparticle interaction we have measured field cooled (FC) $M$ vs $H$ loop for all the nanoparticles. In presence of unidirectional anisotropy, FC hysteresis loop is shifted along the field axis, generally, in opposite direction to the cooling field. This loop shift is commonly known as `Exchange Bias (EB)' which is characterized by field, $H_E$.\cite{nogues} The samples under study have been cooled in +10 kOe from room temperature to 10 K, and after proper thermal stabilization hysteresis loops have been recorded (Fig. 12). We have ensured that the artifacts  due to minor hysteresis loop are eliminated. It is evident in figure that collected $M$ vs $H$ loop is shifted toward the negative field axis. It is worth mentioning that we find similar opposite shift when samples are cooled in negative field. We calculate $H_E$ [= - ($h_{+}$ - $h_{-}$)/2, where $h_{-}$ and $h_{+}$ are the point of intersection on the field axis at decreasing ($-$) and increasing ($+$) fields cycles] as 155.3, 132.4 and 105.7 Oe at 10 K for N600, N650 and N700 sample, respectively. Conventionally, EB is believed to be directly associated with the interface of FM and AF components. However, FM in contact with SG or ferrimagnets are also observed to give rise EB properties.\cite{nogues,mannan} For the present PSMO nanoparticles, we believe that the observed EB is guided by the interparticle interaction rather than the FM/AF interface mechanism. Our conclusion is based on the following points: (i) Fig. 3 conclusively shows the FM nature of low-$T$ magnetic state for all the nanoparticles. (ii) Even if, EB arises due to the presence of residual AF components then $H_E$ would vanish above the FM-AF phase transition temperature $T_N$. We find finite $H_E$ at or above 150 K whereas $T_N$ during cooling in same field occurs around 84 K for bulk PSMO.\cite{psmo-crit} (iii) The residual AF components are supposed to increase with the increasing particle size as the decrease in particle size does not favor AF ordering,\cite{raoncmo,tapati-lcmo} therefore it will result in more FM/AF interfaces. But, our estimated $H_E$ decreases with increasing particle size. (iv) The variation of $H_E$ follows the similar trend of interparticle interaction, i.e., decreases with increasing particle size. These results straightforwardly imply that EB in these PSMO nanoparticles originate due to the mechanism guided by the interparticle interaction. Furthermore, these confirm the unidirectional-anisotropic nature of this interparticle interaction which leads to crossover in critical lines on the H-T plane. Details of EB in these nanoparticles will be published elsewhere. Nonetheless, this interparticle interaction induced crossover in critical lines with fields in these compounds is quite intriguing and requires further studies involving both experimental and theoretical endeavor to comprehend.       

\maketitle\section{Conclusion}
In conclusion, we show that nanoparticles of Pr$_{0.5}$Sr$_{0.5}$MnO$_3$ of three different average sizes have FM ground state contrary to their bulk counterpart in spite of having same crystallographic symmetry. The low temperature FM-AF transition which is a marked feature of bulk compound is significantly absent in nanoparticles.  Detailed linear as well as nonlinear ac susceptibilities coupled with dc magnetization confirm the SPM nature of these nanoparticles. Presence of interactions among these particles has been corroborated from the analysis of frequency dependent peak in ac susceptibility and temperature dependence of coercive field. The nature of this interaction is identified to be of dipolar type and the strength of the interaction is found to decreases with the increase in particle size. This interparticle interaction gives rise to some kind of glassiness in the magnetic response even in these SPM system. The effect of this dipolar interaction on the magnetic properties is clearly evident as the systems exhibit crossover from AT to GT like critical lines with increasing field above their respective interaction field, which is rather intriguing. Following the theoretical prediction, we believe that this crossover phenomenon is induced by the presence of unidirectional anisotropy which arises from the interparticle interaction and the presence of exchange bias effect in these samples confirms it. 
 
\maketitle\section{Acknowledgment}
We are thankful to P. Chaddah for useful discussions. We acknowledge N. P. Lalla for XRD and TEM measurements. We also thank P. Dey for the help in sample preparation and Kranti Kumar for the help in measurements. DST, Government of India is acknowledged for funding VSM. AKP also acknowledges CSIR, India for financial assistance.

\newpage

\begin{figure*}
	\centering
		\includegraphics[width=8cm]{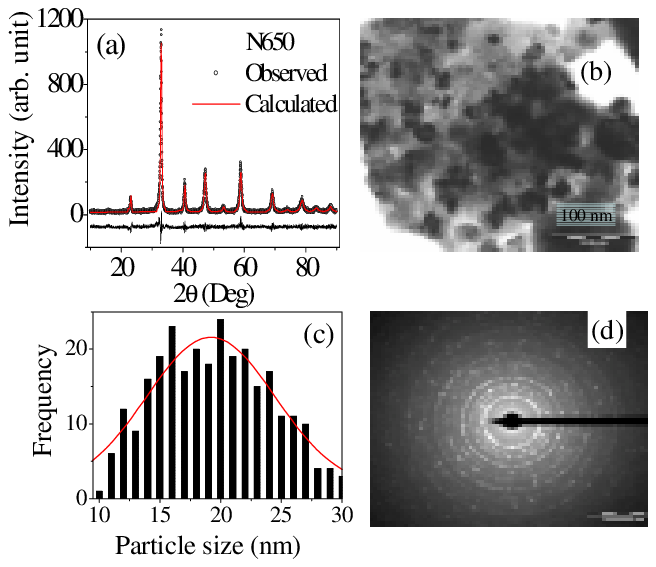}
	\caption{(Color online) (a) Rietveld refinement profile of powder XRD pattern of N650 sample at room temperature. The open circles and solid continuous lines represent the observed and calculated pattern respectively. The difference plot is shown at the bottom of the figure. (b) The bright field TEM image of N650 sample collected at room temperature. (c) Histogram obtained from several TEM images shows particle size distribution for N650 sample. (d) Room temperature selected area electron diffraction (SAED) pattern of N650 sample showing good crystalline nature of nanoparticles.}
	\label{fig:Fig1}
\end{figure*}

\begin{figure*}
	\centering
		\includegraphics[width=8cm]{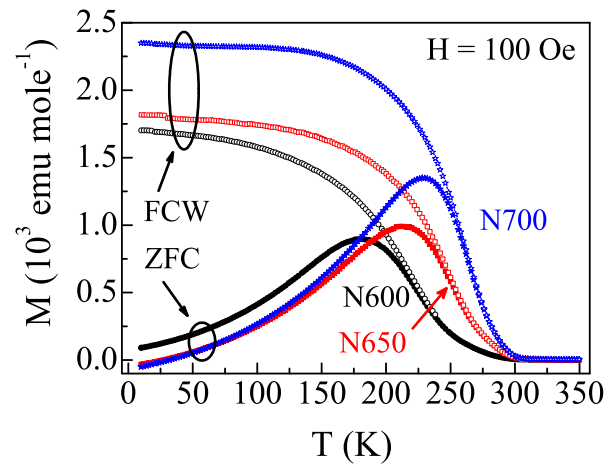}
	\caption{(Color online) Temperature variation in dc magnetization collected in 100 Oe following ZFC and FCW protocols for all the nanoparticles.}
	\label{fig:Fig2}
\end{figure*}
 
\begin{figure*}
	\centering
		\includegraphics[width=8cm]{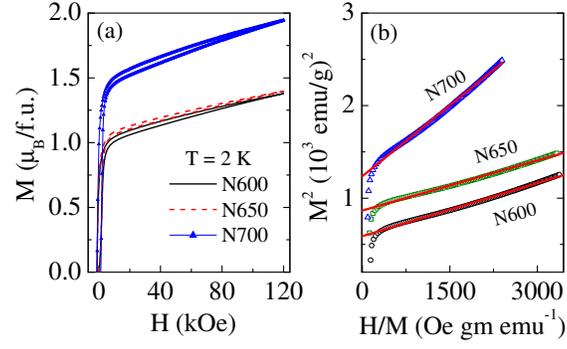}
	\caption{(Color online) (a) Magnetization vs field data are shown for all the nanoparticles at 2 K. (b) Arrott plot ($M^2$ vs $H/M$) of the isotherms in (a) are shown. Lines are due to straight-line fitting of the plots in high fields. The data for N650 sample are vertically shifted by 200 for clarity.}
	\label{fig:Fig3}
\end{figure*}

\begin{figure*}
	\centering
		\includegraphics[width=8cm]{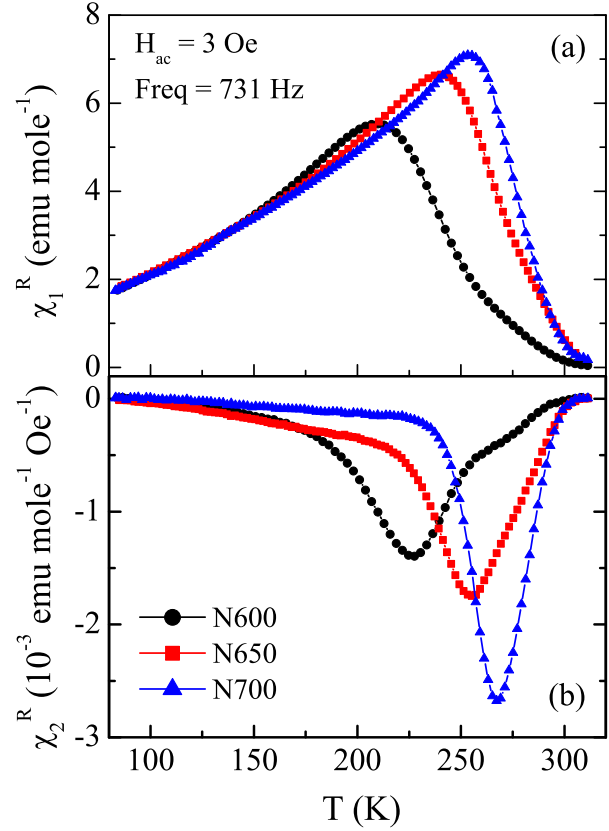}
	\caption{(Color online) (a) Real part of 1st order ac susceptibility measured in field 3 Oe and frequency 731 Hz are plotted as a function of temperature for all the nanoparticles. (b) Real part of 2nd order susceptibility is plotted as a function of temperature for all the nanoparticles in same field and frequency.}
	\label{fig:Fig4}
\end{figure*}

\begin{figure*}
	\centering
		\includegraphics[width=8cm]{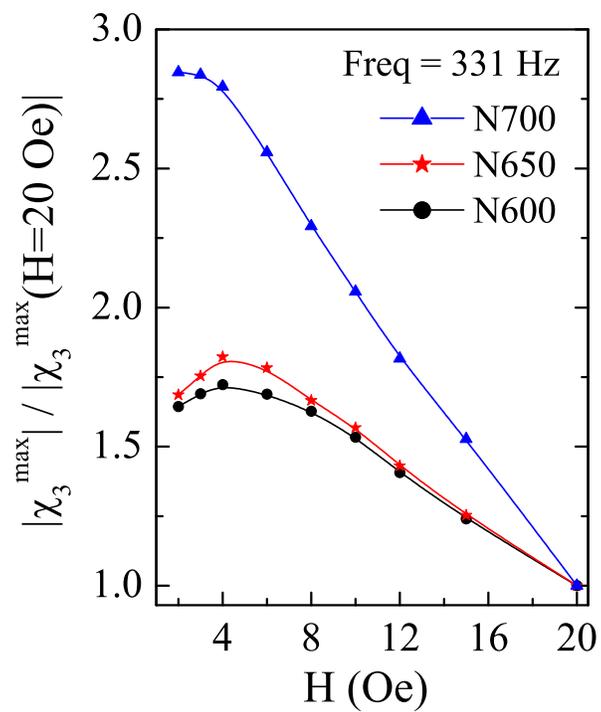}
	\caption{(Color online) The maximum value of $\chi_3$ normalized by that at 20 Oe has been plotted as a function of applied ac field for all the nanoparticles which clearly shows the non-diverging behavior of $\chi_3$ as H $\rightarrow$ 0.}
	\label{fig:Fig5}
\end{figure*}
	
\begin{figure*}
	\centering
		\includegraphics[width=8cm]{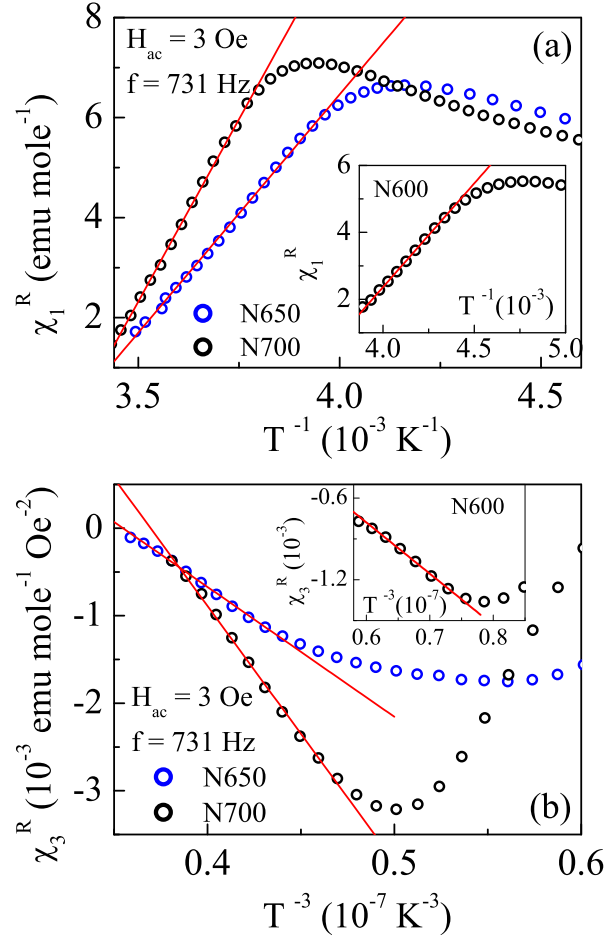}
	\caption{(Color online) Temperature variation in (a) 1st order and (b) 2nd order ac susceptibility is plotted above the blocking temperature for N650 and N700 samples. Straight lines are the (a) T$^{-1}$ fit to $\chi_1$ (Eq. 3) and (b) T$^{-3}$ fit to $\chi_3$ (Eq. 4) data. Inset shows the same plotting for N600 sample in the respective graphs.}
	\label{fig:Fig6}
\end{figure*}

\begin{figure*}
	\centering
		\includegraphics[width=8cm]{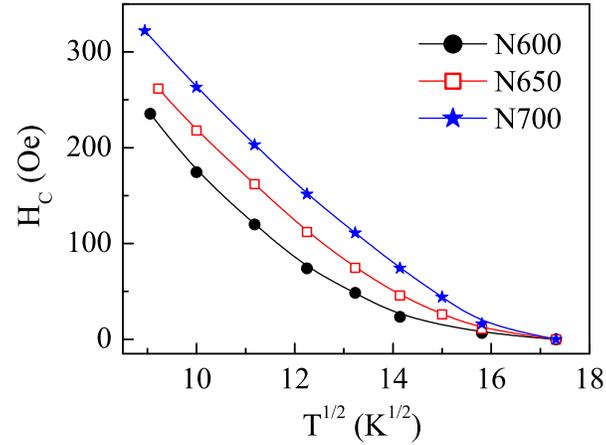}
	\caption{(Color online) Temperature dependence of coercive fields for all the samples.}
	\label{fig:Fig7}
\end{figure*}
	
\begin{figure*}
	\centering
		\includegraphics[width=8cm]{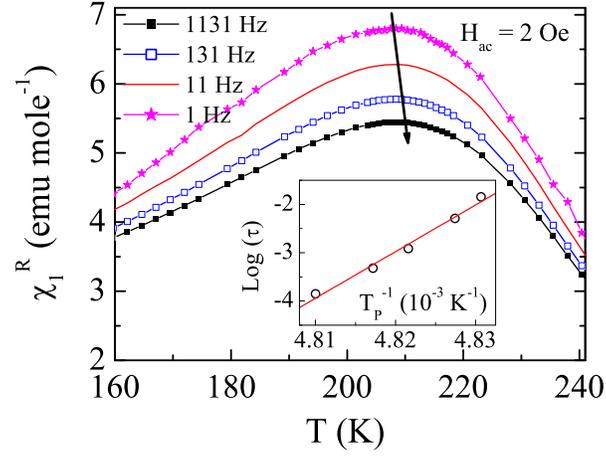}
	\caption{(Color online) Temperature dependence of real part of $\chi_1$ at various frequencies is plotted for N600 compound. Arrow indicates peak shift with the increasing frequencies. Inset shows N$\acute{e}$el-Arrhennius law (Eq. 6) fitting of frequency dependent peak temperature (defined in text).}
	\label{fig:Fig8}
\end{figure*}

\begin{figure*}
	\centering
		\includegraphics[width=8cm]{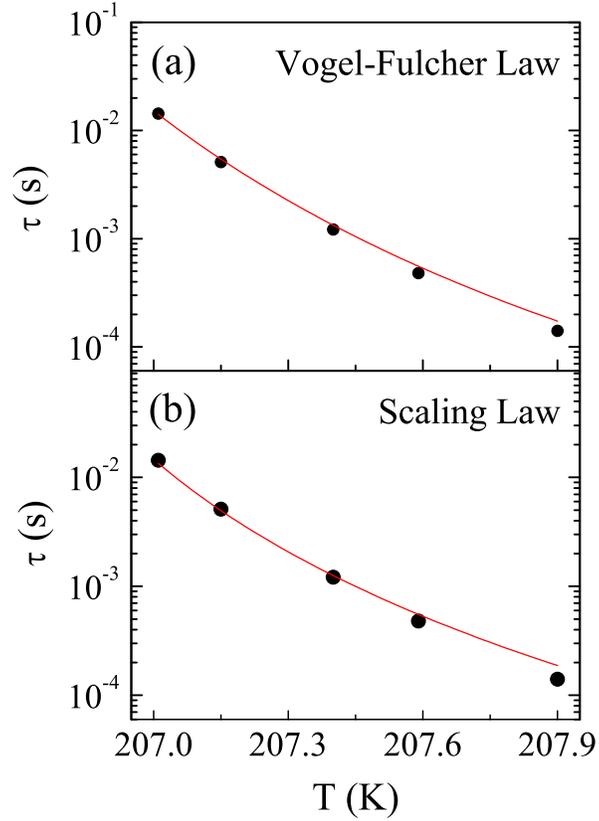}
	\caption{(Color online) The best fit of the relaxation times ($\tau$) to (a) the Vogel-Fulcher law (Eq. 7) and (b) the Scaling law (Eq. 8) for N600 sample.}
	\label{fig:Fig9}
\end{figure*}

\begin{figure*}
	\centering
		\includegraphics[width=8cm]{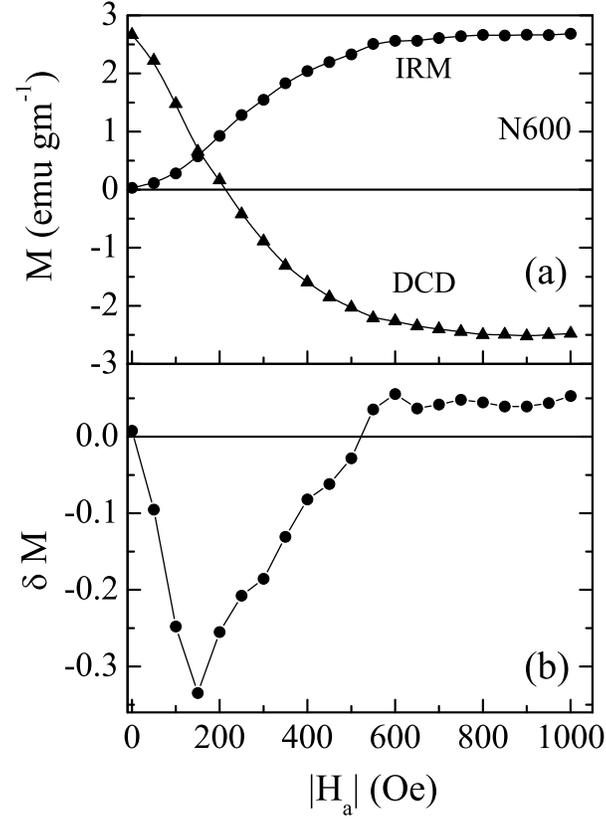}
	\caption{(a) Remnant magnetization measured following IRM and DCD protocols (defined in text) are plotted as a function of applied field for N600 sample. Bottom panel shows calculated $\delta M$ values following Eq. 10 for the same sample.}
	\label{fig:Fig10}
\end{figure*}

\begin{figure*}[h]
	\centering
		\includegraphics[width=8cm]{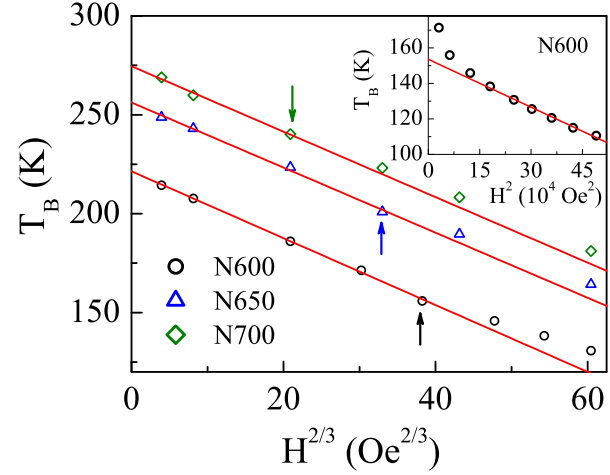}
	\caption{Color online) The temperatures where the peak occurs in ZFC magnetization are plotted as a function of applied field for all the samples. Straight line is fit to AT line (defined in text). The vertical arrows mark the field above which systems deviate from AT line behavior. Inset: Field dependence of peak temperature has been fitted to GT line for N600 sample (defined in text)}
	\label{fig:Fig11}
\end{figure*}

\begin{figure*}[h]
	\centering
		\includegraphics[width=8cm]{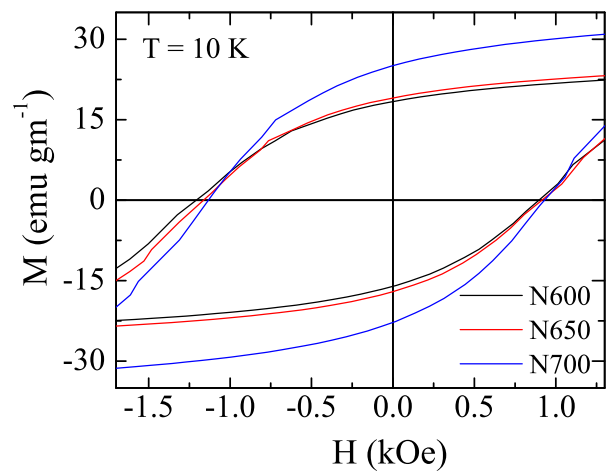}
	\caption{(Color online) Field cooled $M$ vs $H$ loops collected after cooling the samples in 10 kOe from room temperature to 10 K. The recorded loops exhibit shift along the field axis toward negative direction.}
	\label{fig:Fig12}
\end{figure*}

\end{document}